**Information Diffusion, Word-of-mouth effects, and Mutual Funds Performance: A Mathematical Modelling Approach**

Leonard Mushunje

Columbia University, New York, USA

Department of Statistics

lm3748@columbia.edu

## Abstract

This paper puts forward that money managers and prospective investors exchange information and ideas about assets directly through word-of-mouth communication. This, in turn, affects the performance of the mutual funds and their respective managers. We develop a novel "Epi-Finance" model that connects epidemiology and finance to provide a solid explanation behind information transmission in the fund industry. By considering factors such as location, race, gender, and education concerning the equity market, several claims and connections are observed and modeled using an epidemic model. For example, based on location, a mutual fund manager is likelier to buy (or sell) a particular equity at any time if other managers in the $\varepsilon - neighborhoods(\varepsilon > 0)$ are buying (or selling) that same equity. If the information about that particular equity spreads fast and the communication network is well connected, the same pattern shows up even when the fund managers are far apart. Also, a male fund manager is likelier to relate his investment style to other male managers, which is true otherwise. The same fashionable results are expected on race education, among others with special contextual differences.

## 1. Introduction

Shiller (2000) once posited, "…A fundamental observation about human society is that people who communicate regularly think similarly. At some point, there is a Zeitgeist, a spirit of the time…" Word-of-mouth transmission of ideas appears to be a pivotal contributor to day-to-day or hour-to-hour stock market fluctuations." This present study uses an epidemiological modeling approach to explore the value and role of inter-market information transmission through word of mouth. Firstly, we claim that white managers spread market trading information faster than black managers as they conduct more social gatherings, including business meetings and academic conferences than the latter. Also, we hypothesized that close markets share information quickly and experience low volatility with low arbitrage opportunities. Thus, we are interested in the race and distance between the trading points and its effect on information transmission.

Despite the familiarity of the hypothesis about word-of-mouth information transmission, it has yet to receive direct support, with particular attention paid to stock market data. This study extends the hypothesis to mutual funds using a new modeling technique. As in Hong et al. (2005), Shiller and Pound (1989) conducted a survey using a panel of 131 individual investors. They asked them what had drawn their attention to the stock they had most recently purchased.

A fascinating surprise is that most of these investors named a personal contact, such as a friend or relative. Despite this evidence's suggestive nature, the importance of word-of-mouth communication in the financial markets (stock markets, mutual funds, among others) has yet to be fully developed.

This study aims to provide a fair application of the concept in considering mutual funds. Of interest, this paper is part of a small recent literature examining word-of-mouth's effect on financial market operations. The subject appears in Duflo and Saez (2002), Madrian and Shea (2000), Kelly and Grada (2000), and Hong, Kubik, and Stein (2004). Unfortunately, we could hardly find any previous work that has posed the word-of-mouth question in the context of professional money managers- thus the thrust of our study. In light of our opening quote from Shiller (2000), one of the main reasons to be interested in the word-of-mouth phenomenon in the first place is the possibility that it might well explain the market movements and trades in mutual funds and other related executes. Within the objection, individuals are likely to exert a significant influence on stock prices through word-of-mouth communication. We aim to apply the same idea to money management. Despite the narrow exploration of the subject, some related literature exists; Coval and Moskowitz (1999, 2001) analyzed mutual fund managers, where their basic thematic idea was that physical proximity facilitates information transmission, and they focus on the information that investors can gather about nearby companies-an almost similar view to ours. In addition, Hong, Kubik, and Stein (2004) explore the role of social interaction while paying attention to its effects on investment behavior. In their study, they document that more socially active households are also more likely to invest in the stock market as they will have some extra information obtained from social engagements and attention to professional investors and money managers. Argan et al. (n.d) determine the link between word-of-mouth and financial stock behavior. They look at eight banks and their respective customer’s stock transactions in Eskisehir, Turkey. Using Confirmatory factor analysis (CFA), they found that word-of-mouth constructs in the stock exchange could be conceptualized and measured as a three-dimensional construct comprising experimental information, communication skills, and technical expertise.

Huang (2013) explores the causal effect of word-of-mouth communication on investors' trading decisions. Using stock-financed acquisitions as a source of exogenous variation in households' portfolios, they find that in the year after a stock-backed acquisition, both target investors and their neighbors substantially increase their trading intensity in the acquirer industry (excluding the acquirer firm). Nevertheless, no such change is observed after cash-financed acquisitions. This shows the effect and value of communication among the nearby firms. On the same line, Ivkovich and Weisbenner (2007) find that when investors purchase a stock from a particular industry, other neighboring investors from the same industry increase their stock purchases. Mushunje and Mashasha (2020) looked at the Gompertz model, which shares some properties with the SIR model employed in this study. They applied the model to measure and assess the growth behavior and development of the insurance industry in South Africa. While Cohen, Frazzini, and Malloy (2008) look at the transmission channels that consider investors and firms, Cohen, Frazzini, and Malloy (2010) explore the one between investors and financial analysts, and Engelberg, Gao, and Parsons (2010) consider banks and firms.

All these studies explain the value of inter-agent communication and its effect on market trades. Hellstrom et al. (2013) find new empirical evidence on the impact of family on individuals' stock market decision-making and participation. Their main results indicate that an increase (decrease) in individuals' likelihood for subsequent participation is well explained by positive (negative) parental and partner stock market experiences. Of course, Hong et al. (2005) and other literature mentioned above failed to establish a clear link between word-of-mouth communication and stock prices, and their findings left the door open wider than before. This motivated us to extend the idea to mutual funds while introducing a new modeling technique of the phenomena.

## 2. Methodology

### 2.1.Data

We extracted our data from various spectrums. We consulted the monster databases and the mutual funds' archives at JSE in South Africa. We conveniently performed data augmentations to meet our hypothesis tests in three ways. Firstly, we associate each mutual fund with a fund family and use the Fund Index-Directory of Investment Managers to establish the province where the fund family is headquartered from 2010 to 2018. Secondly, we use the Disclosure database for each stock to find the location of the company's headquarters. Lastly, we use each firm's executive records to identify the managers' race (black or white) over the targeted period. We then analyze both funds' holdings and their trading behavior, primarily emphasizing trade-based specifications while incorporating the effects of race and location. We obtained our final sample of the number of mutual funds in stages. Firstly, we only consider those funds whose headquarters are in the provinces of South Africa and for whom a family location is identifiable. Secondly, we eliminate funds not influenced by word of mouth (all index funds), since their behavior is purely mechanically determined. Thirdly, stock mutual funds are only considered. Finally, we proxy to consider firms that report semi-annually and annually to consider all mutual funds in the Spectrum databases.

We eliminate funds that do not have the same management company as the plurality of other funds in the same family for compatibility and good statistical sounds. This is based on the inference that these funds are contracted out and may occur in a different province than the family's headquarters. Since location is one of the critical variables in our study, such funds can lead to measurement errors and biased results. Our final screened funds sample data comprises 200 funds as of December 2018.

## 2.2.Model

Before introducing our primary proposed epidemic model, we shall provide two distinct modified gossiping models relevant to information transfer within mutual fund markets. (i). A pair-wise information exchange model shows a schematic and state transfer of information within nearby markets. (ii). We used an information diffusivity model between fund families and from market to market. The idea is to capture and represent information transmission among the managers, considering the two main factors (race and location). Clear mathematical structures of the models are subsequently provided below:

### 2.2.1. A Markovian pair-wise information exchange model

We inherit the gossip protocol model Demers et al. developed (1987). We modified the model by inducing the Markovian transition rates. The transition rates measure either the information gain or information loss with effects. Since our study goal is to explore the effects of information exchange between fund managers, it is necessary to consider the transfer intensities (rates). The essence of our model is that considering all the fund families in each South African province, we group the families pair-wise over the market as our geometrical surface. From the model, transition rates are identified, and so is the information loss or gain. The effects of these end-point results (gains/losses) are then expressed in terms of the fund performance and arbitrage analysis. Fund families/types whose performance is independent and identically (IID) are considered to have resulted from information exchange gain and true otherwise. The reader is referred to the results section for better understanding.

### 2.2.2. Model assumptions

We assume there are connected communication links between the fund managers in different provinces. This leads to a further assumption that no information transfer restrictions exist. The managers can exchange information whenever they want but generally under pair-wise conditions. Further, we assume that every message sent is positively likely lost or gained due to communication medium constraint status.

Moreover, message losses or gains occur with the same probability and are independent. Lastly, we assume that when a manager sends a message, he needs to know whether the message is delivered successfully and neither its effects. We, thus, give a scenario representation of the information gains and losses in a pair-wise fashion as below:

Figure1. Status Scenarios of Information Exchange

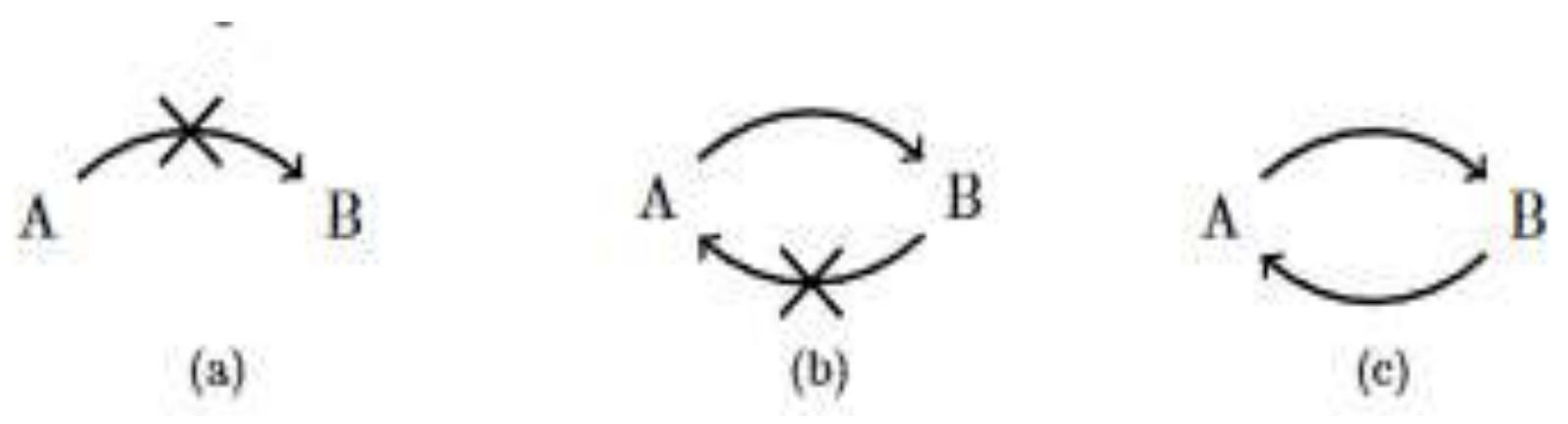

We present the information exchange scenarios status between two connected managers in a connected social network. We present three cases with lossy information routes: (a) loss of information from the sender, (b) loss of feedback from the receiver, and (c) complete information exchange network (no loss observed).

The model with Markovian rates is given in Figure 2 below. Using the model, we analyze the spread of investment information $d$ in a network of fund managers executing different money investment ventures. In this analysis, we assume that the communication channels are prone to both gains and losses. The model has four possible states of the caches of A and B before and after the information transition: The corresponding states are (0,0), (0,1), (1,0), and (1,1) in the state transition model in Figure 2. Note that the states refer to different fund families in different fund markets with different fund managers. The model is developed under the Markovian conditions of state relations. Unlike in information models without failures, as in Bakhshi (2011), in our model, the state (0,0) is no longer isolated since there is a possibility of information gain between the states. Transitions from one state to another are labeled by the respective transition probabilities $p(A_2B_2|A_1B_1)$, where $A_1B_1$ Is the state before an exchange, and $A_2B_2$ Is the state after the exchange, with $A_I, B_I \in \{0,1\}$. The bits $A_1, A_2$ And $B_1, B_2$ Indicate the presence (if equal to 1) or the absence (if equal to 0) of a generic item $d$ in the cache of an initiator manager A, and the contacted manager B, respectively.

To compute our Markovian transition probabilities, we express all transition probabilities, $P(A_2\ B_2\ |A_1B_1)$, of the state diagram in terms of four probabilities as follows: 1. Probability Pselect of a message to be selected by another manager during an exchange, 2. Probability: Pdrop of a message likely to be dropped during an exchange/conversation. 3. Probability Ploss that a message is not delivered to another manager due to connection breakdown, and Pgain that a message will reach the intended manager. Note that this study uses the time-homogenous model, that is, the time-independent model. Of course, we are interested in effective information transfer and exchange and all the linked information arbitrages, but the messages' duration does matter. However, we will use a time-homogenous model for simplicity. Thus, our study prepares another way of modeling where a time inhomogeneous model can be considered. For notation, we write P¬select for 1 − Pselect, P¬drop for 1 − Pdrop, P¬loss for 1 – Ploss, and P¬gain for 1 – Pgain.

**Figure2: The pair-wise information transition model**

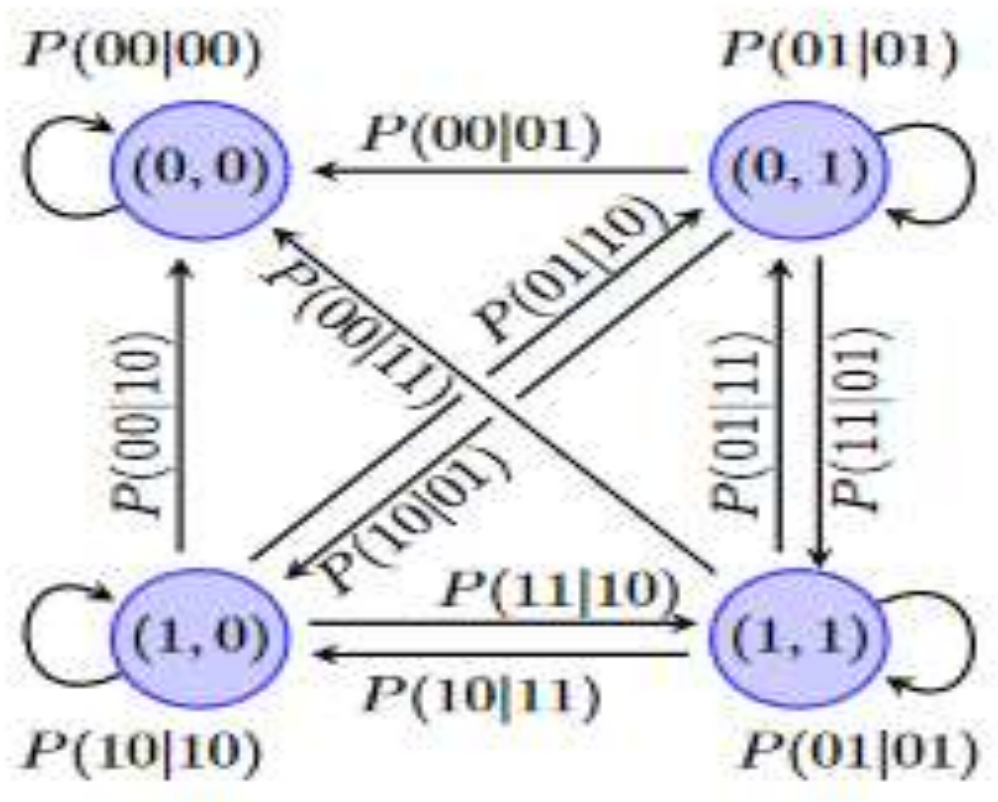

In the subsequent results section, we give some simple illustrations with associated probabilities for our pair-wise model. We now resort to our second gossiping model of information diffusivity.

### 2.2.3. Information Diffusivity model

We must consider the intensity and extensity of the flow of information from market to market, fund family, A to fund family, B or money manager A to money manager B, whichever way. We employed the model for network communication presented by Banerjee et al. (2019) to model our phenomenon of investment market information spread. The managerial style paper postulates that communication between fund managers has some notable effects on investment behavior and market performance. Information spread and diffusion are thus worthwhile modeling. The model is formulated as follows:

We are considering a network of investors consisting of n individuals connected via a randomly defined network with an adjacency matrix $\mathbf{w} \in [0,1]^{N\times n}$. The i th entry is the relative probability with which the investor says something to the $J^{th}$ Investor. However, this relation needs not be reciprocal. Otherwise, there might be information loss, which leads to model failure. We take the network **w** to be random and variable and let $V^{(R,1)}$ Be its first (right-hand) eigenvector, corresponding to the largest eigenvalue$\Lambda_1$. Following the Perron–Frobenius Theorem, the first eigenvector is nonnegative and real-valued. Throughout what follows, we assume that the network is (firmly) connected in that there exists a random path from every manager/investor to every other investor so that information coming from any investor could eventually make its way to any other investor. When modeling word-of-mouth effects on any phenomenon, the concepts of diffusion centrality and network gossip are central and, thus, should be factored in.

### 2.2.4. Diffusion centrality

The notion of diffusion centrality is well explained in Banerjee, Chandrasekhar, Duflo, and Jackson (2013). They posited that the concept is based on the random flow of information through a network, based on a process that underlies many contagion models. Relatedly, this study conjectured that information randomly diffuses from one manager to the next through the contagion space. The random algorithm operates as follows;

Initially, a piece of information about the asset or market itself is initiated at manager i and then broadcast outwards from that manager. For a given time-space, **S**, with probability $W_{Ij} \in [0,1]$, independently across pairs of neighbors, each informed manager $i$ informs each of its neighbors $j$ of the piece of information about the asset in question. In most cases of financial markets, the information exchange process operates for a finitely defined **T** periods, where $\in$ $\mathbf{Z}^+$. T is typically finite because new market information may only be relevant for a limited time. In some cases, new information can lead to a change in conversation between managers.

Usually, information spread is directly related to the market topics. As such, if the market topic changes, the new information being spread may no longer be considered, and it may be dropped. According to Banerjee et al. (2013), information diffusion centrality measures how extensively the information spreads and is expressed as a function of the initial point (manager). More precisely, let us define the inception function for the targeted manager using the hearing matrix as flows:

$\mathbf{H}(\mathbf{w}, \mathrm{T}) = \sum_{\mathrm{T}=1}^{\mathrm{T}} (\mathrm{W})^{\mathrm{T}}$ ,,,,,,,,,,,, (1). The $ij_{th}$ entry of $\mathbf{H}$, $\mathrm{H}(\mathbf{w}, \mathrm{T})_{\mathrm{Ij}}$ is the expected number of times, within T periods, that the j manager hears about a piece of information originating from $the\ i$ manager. Finally, diffusion centrality is then defined as follows:

$$\mathrm{DC}(\mathbf{w}, \mathrm{T}) = \mathbf{H}(\mathbf{w}, \mathrm{T}) \cdot \mathbf{1} = \sum_{\mathrm{T}=1}^{\mathrm{T}} (\mathrm{W})^{\mathrm{T}} . \mathbf{1} ,,,,,,,,,,,,, (\mathbf{2})$$

The reader is urged to consult Jackson and Yariv (2005), Bloch and Tebaldi (2016), Jackson (2013), Lawyer (2014), and Banerjee et al. (2019) for background and other models of diffusion and contagion as it is outside the scope of this text. Therefore, $\mathrm{DC}(\mathbf{w}, \mathrm{T}\,)_{\mathrm{I}}$ Measures the expected total number of times that some information that originates from $the\ i$ manager is heard by any other manager in the fund industry during a finite T –period time interval. The time taken for the information to diffuse from one manager to the next is termed information extensiveness, and the possibility of some feedback effects is termed information intensity. These two ideas are important when measuring and modeling the word-of-mouth effects in financial markets.

We provided the two main information exchange models in the context of fund managers in the fund industry, assuming a purely contagion environment. Precisely quoting, like the spread of the recent pandemic of COVID-19 through human interaction and contagiousness, news, information, and media about financial assets and market information can be spread likewise as most markets have no barriers to entry and exit. We propose a related model based on the widely epidemic-used SIR model. The model is explained as follows:

### 2.2.5. Epi-finance model- Epidemiology in finance- A selected SIR model

Our paper is centered on modeling the information transfers within mutual funds. Our hypothesis to be tested is based on the fact that information is spread and transmitted through word of mouth, like in epidemiology. Following that, we shall adopt and apply the SIR model. The application of the model, which implies an interdisciplinary application, is new and novel. In this case, we consider any interesting and valuable information a disease that can be transmitted between the investors and fund managers within the active markets. When modeling such transmissions, we are considering the population size of the possible information sources and transmitters and the speed of the information flow. This further means that our model captures the information flow rate through the fast and slow dynamics diffusion component.

#### 2.2.5.1.The SIR-finance model

We follow the SIR model that Kermack and McKendrick proposed (1932). The model has three main components: i). susceptible class, which is proportional to the population size, is the number of fund managers under study. Ideally, every manager is susceptible to information transmission and the associated influence. ii). The infectious group measures the total number of managers with negative and positive information to transmit to other managers, both in-city and outside the manager's own cities. iii). The Recovery group consists of the managers who have once received negative information that could have forced them to close or to operate uneconomically and have recovered from these information misleads.

The underlying hypothesis in the model is that, in the presence of negative information content, managers tend to underperform and possibly close their funds (especially the so-called toddler managers). On the other hand, if the information in transit is positive, the market setup and manager's performance boosts incredibly. The two phenomena underpinning our model are represented by $\alpha \; and \; \mu$, respectively. If these parameters are equal, there is a balance in the system in that no information-induced changes are realized. Our model is bounded in size based on the ideal proposed by Hethcote (1976). We kept our number to N, where there are barriers to entry and exit of managers, except for recoveries (re-entry). The model is given as:

$S' = -\beta SI + \mu(N - S),,,,,,,,,,,,,,(3)$

$I' = \beta SI - \alpha I - \mu I,,,,,,,,,,,,,,,,,,,,(4)$

$R' = \alpha I - \mu R,,,,,,,,,,,,,,,,,,,,,,,,,,(5)$

In this case, $S + I + R = N \; and \; N = 0$. $N$ is constant, and $R$ is obtained when $S$ and $I$ are known. By considering a two-dimensional system of S and I, we have the following set of equations:

$S' = -\beta SI + \mu(N - S),,,,,,,,,,,,,,(6)$

$I' = \beta SI - \alpha I - \mu I,,,,,,,,,,,,,,,,,,,(7)$

Critically, the total number of managers varies with time due to market imperfections (free entry and exit). Thus, using N(t)—the following schematic diagram for information flow from the susceptible group to the recovery class is reasonable.

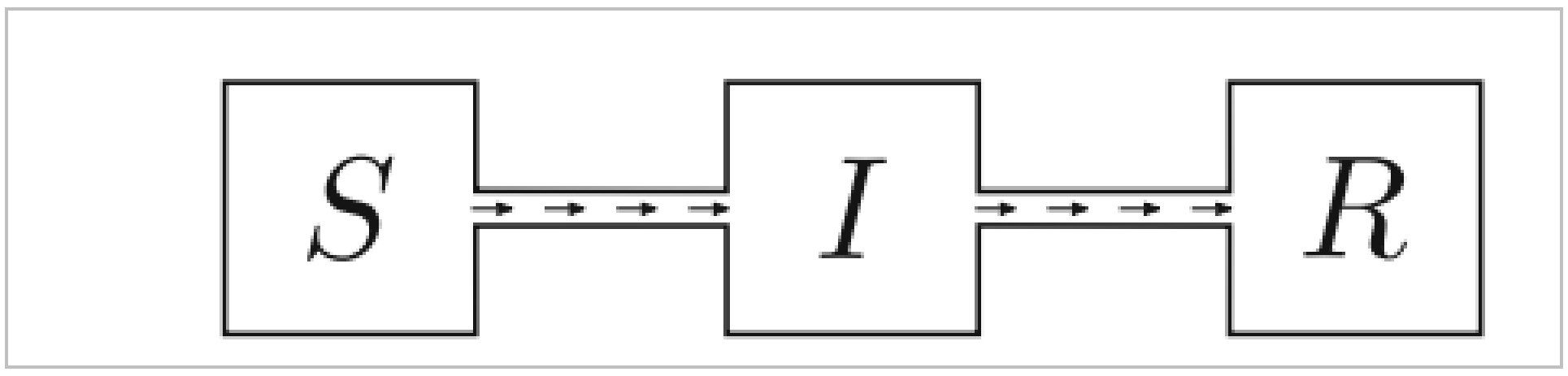

Figure 3: Schematic diagram showing the information flow from one class to the other.

The diagram shows that market information can be transmitted from one class to the next in the same fashion, followed by contagious diseases. Word of mouth can transmit information from one manager group to the next. Below, we provide further analytics and arguments.

**2.2.5.2. Singular perturbations for the response rates to the transmitted information**

We are interested in analyzing the response rates of the investors who are incident to the transmitted information within the market. We propose a mathematical-based method related to the differential equations in use. Singular perturbation is employed to comprehensively analyze the notion based on multiplying the derivative from the differential equation by a parameter whose values can be set until zero. Market information flows in different volumes and at different rates, depending mainly on the nature of the markets and the players involved, among other factors. This then leads to a situation where some investors get information earlier than others, which results in some fundamental trading theorems like the Arbitrage theory. To thoroughly analyze this notion mathematically, we shall employ the Singular perturbation method as below.

**2.2.5.3.Singular Perturbations**

Theoretically, the case of Singular perturbation is commonly identified in systems of differential equations with a small parameter. Consider the following system:

$$\frac{Dy}{D\tau} = f(y, z, ), y(0) = Y_0, , , , , , , , , , , , , , , , , (8)$$

$$\frac{Dz}{D\tau} = g(y, z, ), z(0) = Z_0, , , , , , , , , , , , , , , , , , (9)$$

The above system is solved to give the solution as: $(y(\tau, ), z(\tau, ))$. If we set the system equal to zero, we get a more reduced form as below:

$$F(y, z, 0) = 0, , , , , , , , , , , , , , , , , (10)$$

$\frac{Dz}{D\tau} = g(y, z, 0), z(0) = Z_0, , , , , , , , , , , , , , , , , (11)$ where $(y_0(\tau), z_0(\tau))$ is the solution.

**Notes**

The model parameter is $\tau$. Our perturbation allows only the parameter to be small. The perturbation results imply that the y reaction time will be much faster than the z reaction time. This means y stabilizes faster and reaches the equilibrium point $f(y, z, 0) = 0$ more quickly than the latter. Consider y as an investor from a market in Gauteng province and z as the one from Northern Cape provinces. By varying the model parameter above, the rate at which information and market news reach out to these investors is genuinely different and matters. The first one to receive the information is most likely to act. This is fundamental within the financial trading circles. So, in financial implications, the theory of perturbations is considerable on the rate of information and news flow in the market from investor A to investor B. This is best solved and presented using the reduced form equation above.

#### 2.2.5.4. Special case: Heterogeneous fund markets and information flows

In most cases, markets are inhomogeneous. This makes up the point that markets and the related news and information are non-linear. This implies that making some insights into investors' response behavior to non-linear information is fundamental to both existing and potential investors, including the fund managers themselves. We put forward that market news is unpredictable and thus not linear. We provide a non-linear reaction-diffusion model for such a case as:

$U_T(x,t) = DU_{xx}(x,t) + g(u),,,,,,,,,,,,,(12)$. Here, $g(u)$ is a non-linear flow rate of information with $g(0) = g(K) = 0, g(u) > 0$ for $0 \leq u < K$ and $g(K) < 0$. The flow rate is a function of several factors, such as information inducers, carriers, and users. The differential model can be solved by several means, including the reduction approach, where it can be reduced to the form:

$U = g(u)$. The model is better analyzed using the Jacobian matrix approach, enabling us to analyze severally derived equilibrium points in detail. These points are insightful. Contextually, the points provide an optimal point where fair news and information flows are achieved, which results in market efficiency with no arbitrage opportunities.

## 3. Results and Findings

This section provides some basic statistics of our data components.

**Summary Statistics**

| Sub Category | Mean | Standard deviation | Minimum | Maximum |
|---|---|---|---|---|
| Top 200 stocks, eight provinces | 0.310 | 0.750 | 0 | 200 |
| Simple specification | 0.209 | 0.076 | 0 | 150 |
| Excluding large families (20% of a city) | 0.416 | 0.509 | 1 | 135 |
| Excluding local stocks | 0.227 | 0.480 | 0 | 120 |
| Extensive margin | 0.311 | 0.987 | 1 | 90 |
| Intensive margin only | 0.746 | 0.452 | 1 | 70 |
| Low book-to-market stocks | 0.338 | 0.742 | 0 | 170 |
| High book-to-market stocks | 0.103 | 0.356 | 0 | 60 |

Table 1 gives important statistical properties such as the mean, standard deviation, maximum, and minimum values of the sampled funds data used in this study. We first grouped the funds into eight categories: performance ratios. Funds from the Intensive margin-only group have a significant mean value of 0.746, while the high book-to-market stocks category retained the lowest mean value of 0.103. We find more variation under the extensive margin group; the rest portrayed moderate variations. Grouping enabled us to robustly deal with outliers, hence more stable and insightful results.

**Table 2: Funds Composition by Province**

| | Province | Number of Fund Families | Number Of Funds | % of All Funds | % of All Fund Assets |
|---|---|---|---|---|---|
| 1 | Gauteng | 106 | 211 | 17.8 | 11.9 |
| 2 | N.Cape | 32 | 24 | 10.8 | 9.1 |
| 3 | W.Cape | 24 | 51 | 16.0 | 5.4 |
| 4 | KZN | 33 | 75 | 7.8 | 7.3 |
| 5 | Limpopo | 17 | 34 | 3.1 | 0.3 |
| 6 | North West | 46 | 61 | 5.9 | 1.5 |
| 7 | Mpungalanga | 18 | 53 | 3.7 | 1.5 |
| 8 | Free State | 20 | 47 | 3.1 | 2.8 |

Table 2: Provinces ranked by Number of Fund Families, December 2018. The Table gives basic information on the funds' cross-provincial status and their respective fund families. We rank provinces by the number of fund families in each. We also tabulate the number of individual funds in each province and each province's share of the mutual fund market, stated relative to the total number of funds in our sample and the dollar value of all fund assets.

**Table3. Managed funds and the respective managers**

| | Fund Category | Fund' Managers Race | Fund Mangers' Gender | % of All managed Funds | % of All Fund Assets |
|---|---|---|---|---|---|
| 1 | A | White | M | 17.8 | 11.9 |
| 2 | B | White | M | 10.8 | 9.1 |
| 3 | C | Black | M | 16.0 | 5.4 |
| 4 | D | White | F | 7.8 | 7.3 |
| 5 | E | White | M | 3.1 | 0.3 |
| 6 | F | White | F | 5.9 | 1.5 |
| 7 | G | Black | F | 3.7 | 1.5 |
| 8 | H | White | M | 3.1 | 2.8 |

To further our study, we incorporated the manager's composition by demographics. We provide salient information about managed funds and the respective managers for each fund family with specified gender and race. We note that most fund managers in South Africa are male and white. We interestingly draw out that even though black managers are underrepresented, they perform better than their white counterparts in some instances. This is an interesting result that needs special attention in the future.

### 3.1.Simulations results

We present the model simulation results with different parameter values. The model results show how information is transferred and transmitted from investor to investor within the market social networks.

**Figure 4: SIR information transmission model.**

We adopted and modified the SIR Model to simulate information transmission within a connected market network over time. The model consists of a system of three differential equations that express the rates of change of three variables over time. The three variables are S, I, and R. The model is applied to how information is transferred from one participant to another and the overall market's effect. Analyzing such evolutions is vital in making trading decisions.

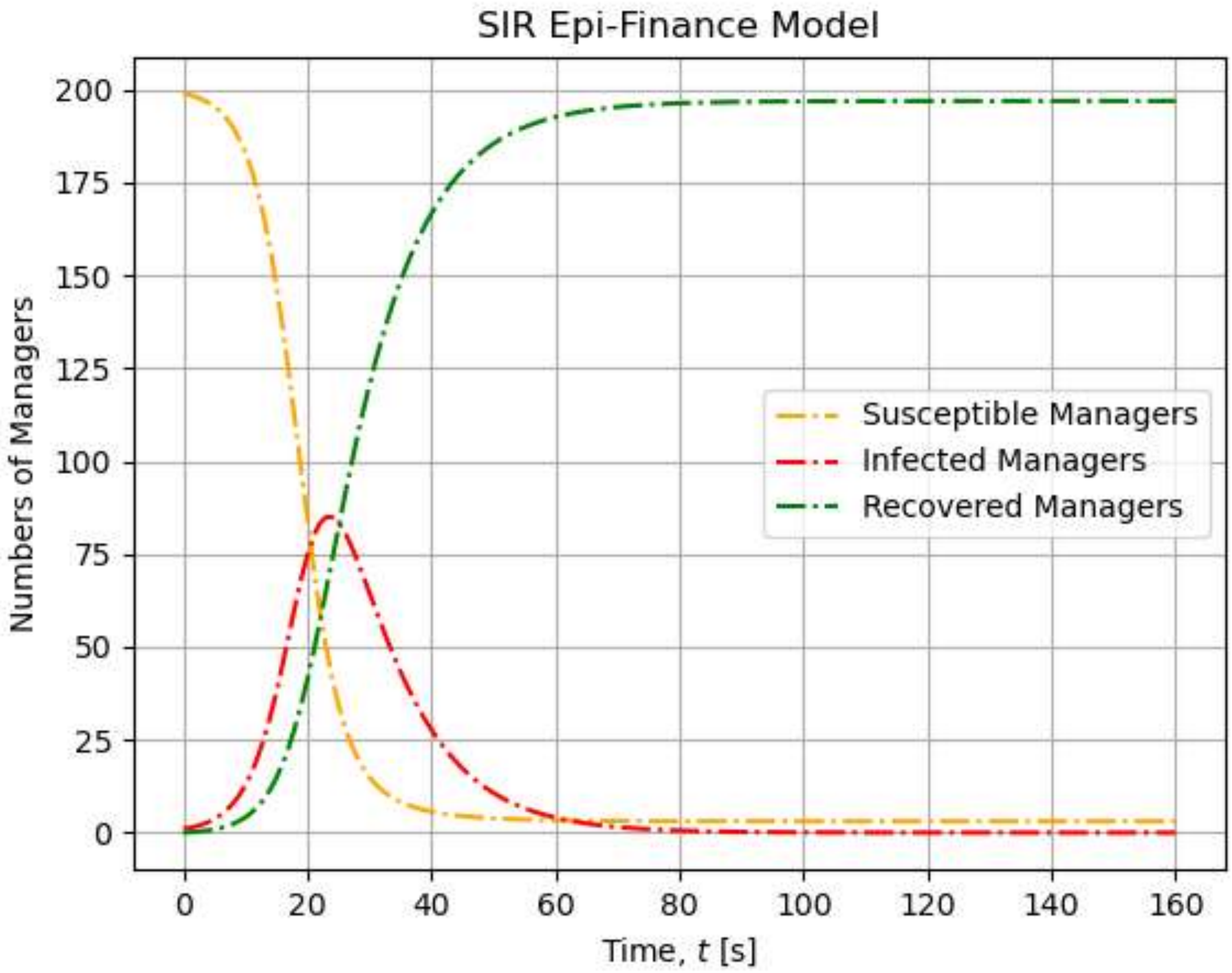

**Notes:** S are the susceptible money managers who are exposed to receiving the market information. I- are the managers with information already capable of spreading the information to other participants. R is the managers who received the information (news) but were not affected by the negative effect of the news. The results show that the number of participants affected by insider information is an increasing function of time up to a certain point. Beyond a certain mark-time, the number starts to fall, and in the long run, market traders become resistant to any transmitted information from other players. This is attributed to factors like deep market experience and stabilization (equilibrium reading). We also present separate simulation results for the infected managers to infer their behavior as more news flows over time in Figure 2.

**Figure 5: Infected managers (Information recipients)**

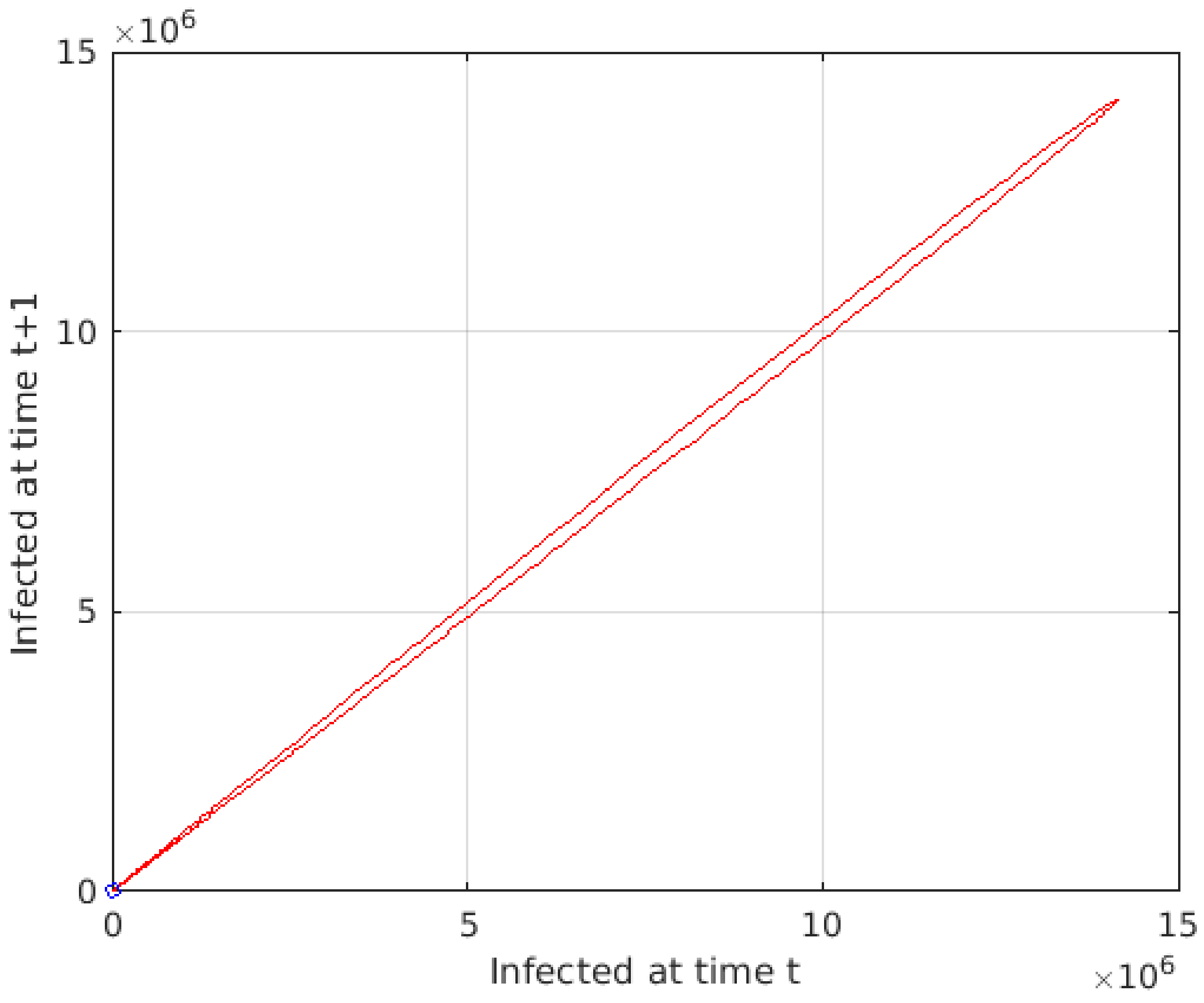

With time, the number of managers affected negatively by insider and market outsider information is a closed loop. It implies that the number increases to a certain threshold, and beyond that maximum point, the number returns to the initial point. The maximum point varies by market size, location, and other factors.

**Figure 6: Singular Perturbation results**

We present the behavior of the SIR model results above given different sets of conditions for the market flow hypothesis resistance from different managers. We aim to see how the participants incept news/information as time passes. In our model, beta and gamma are the hyper-parameters representing the market flow (transmission rate) and reception (response rate). We obtain different relationships and information flow schemes for different parameter values.

Figure 6a): General singular perturbations (Beta = 0, gamma = 0).

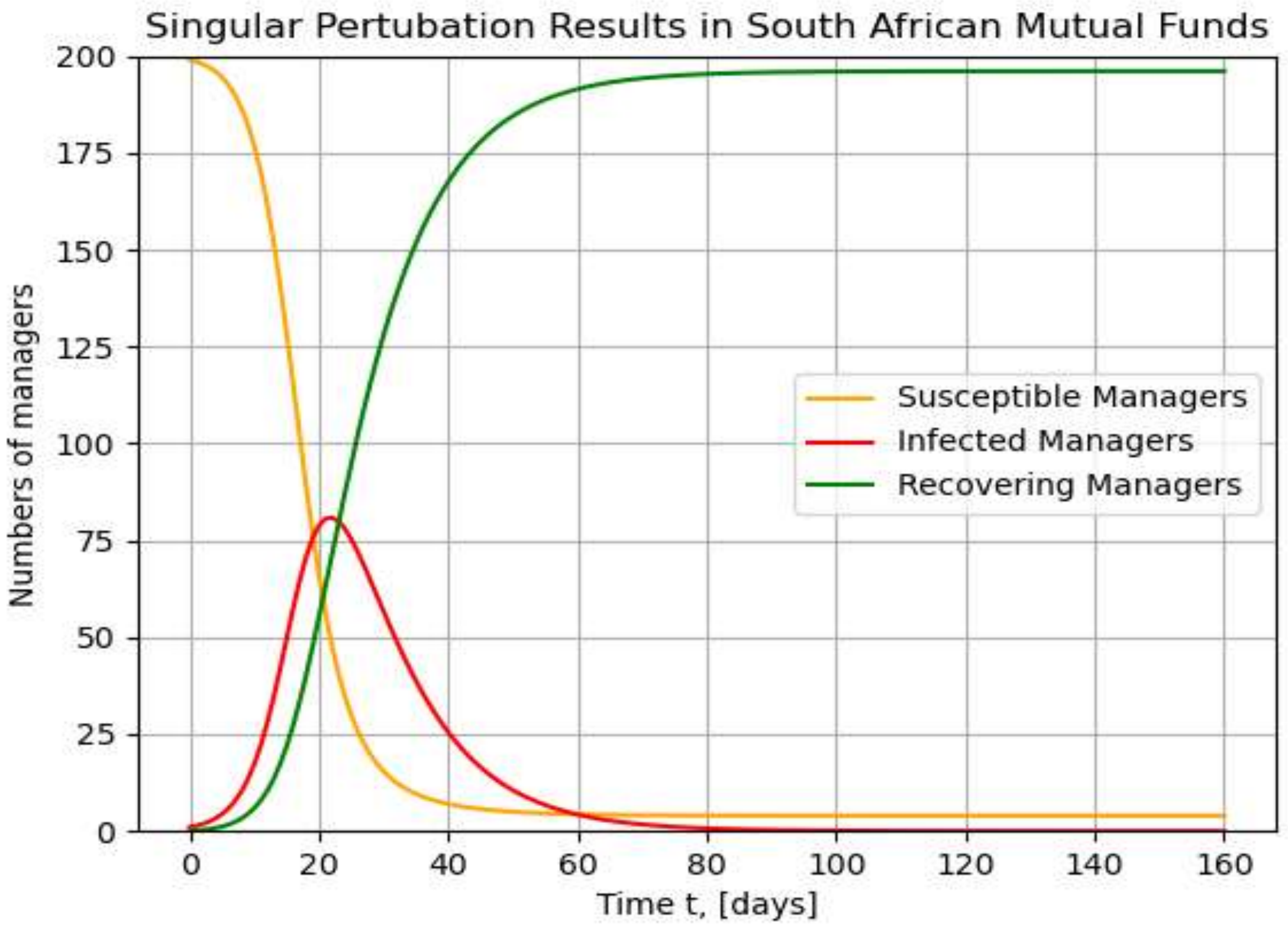

Figure 6b: Simulations for beta = 0.20, and gamma = 0.10

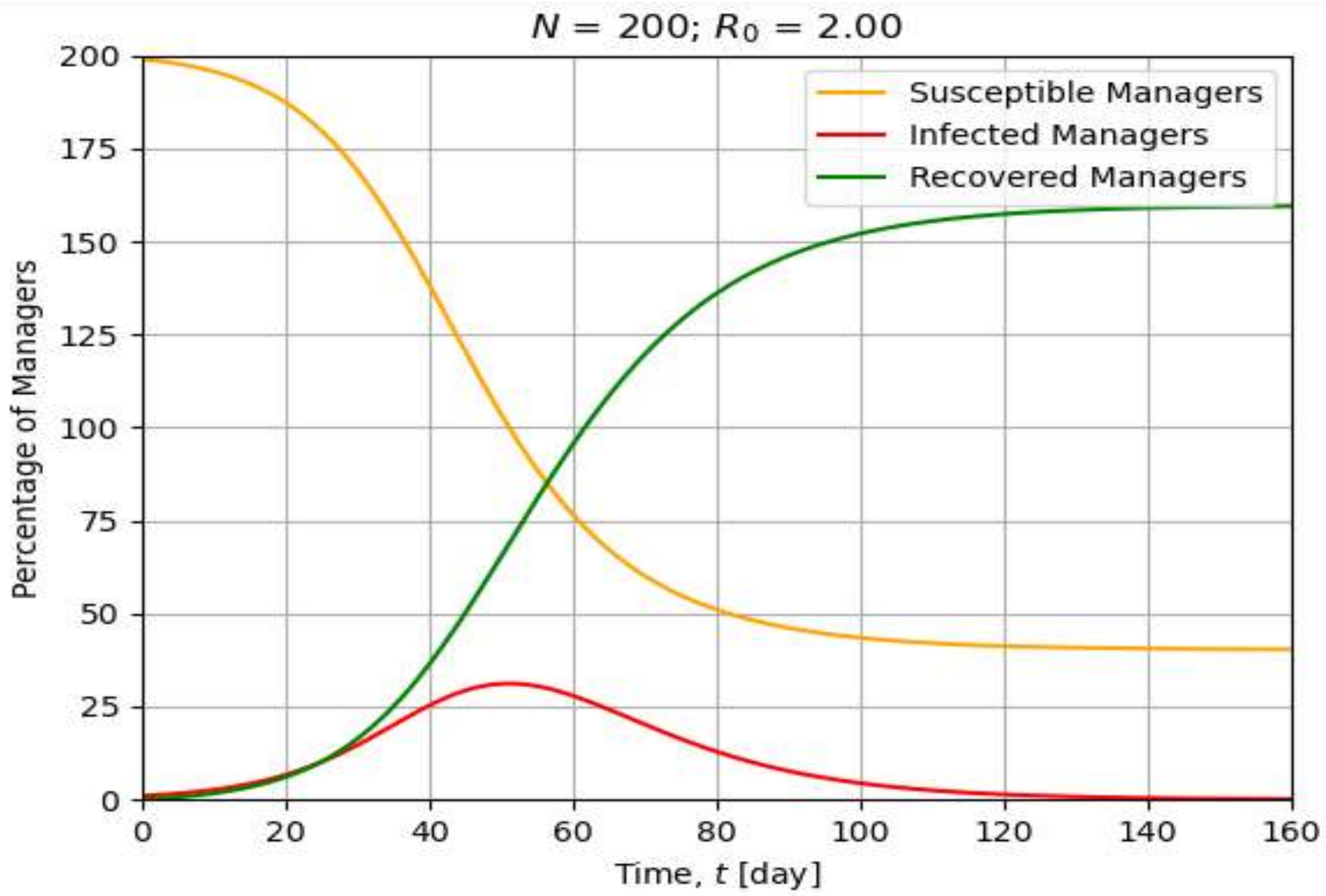

Figure 6c): Simulations for beta = 0.82, and gamma = 0.18

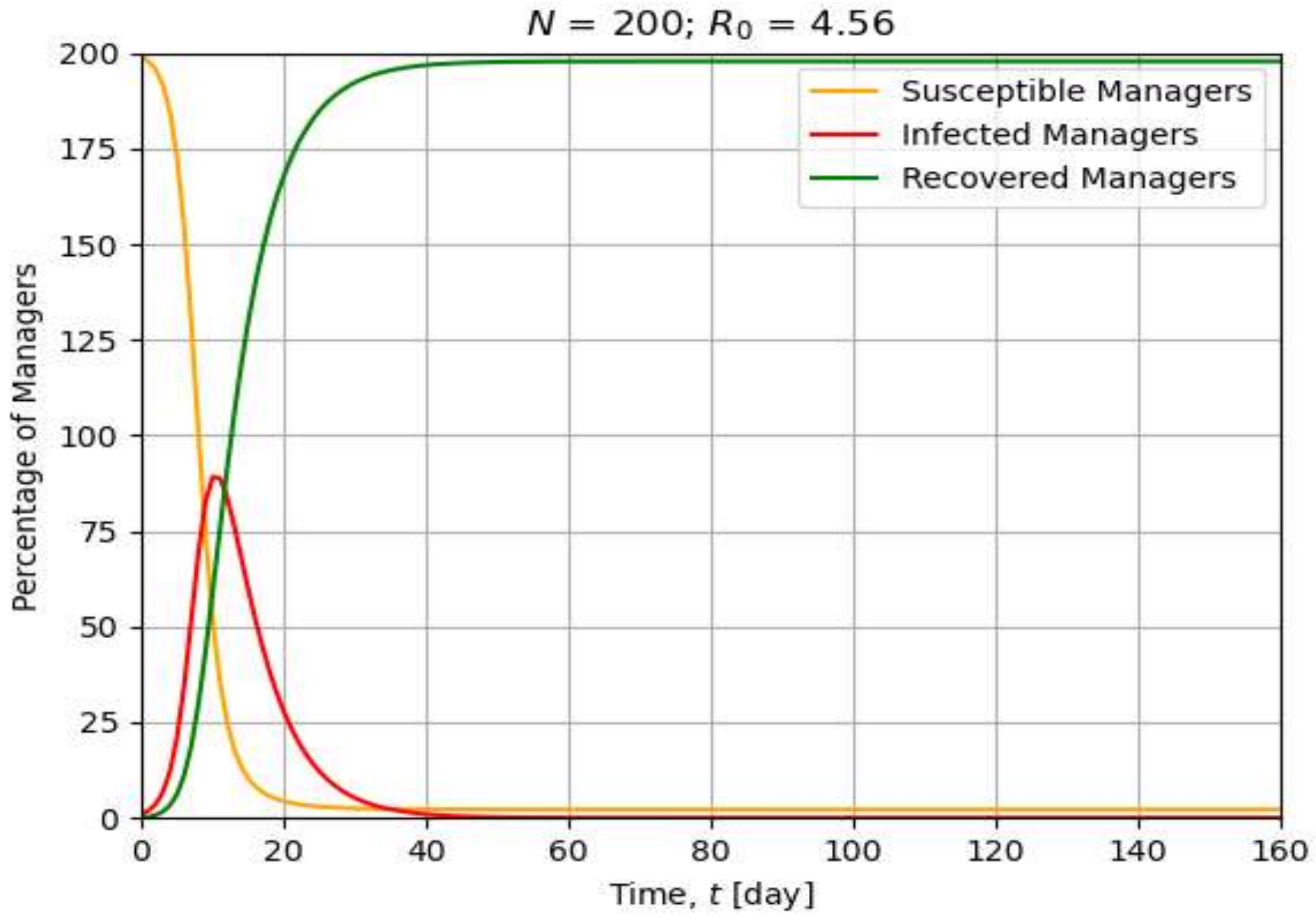

Figure 6d): Simulations for beta = 0.61, and gamma = 0.49

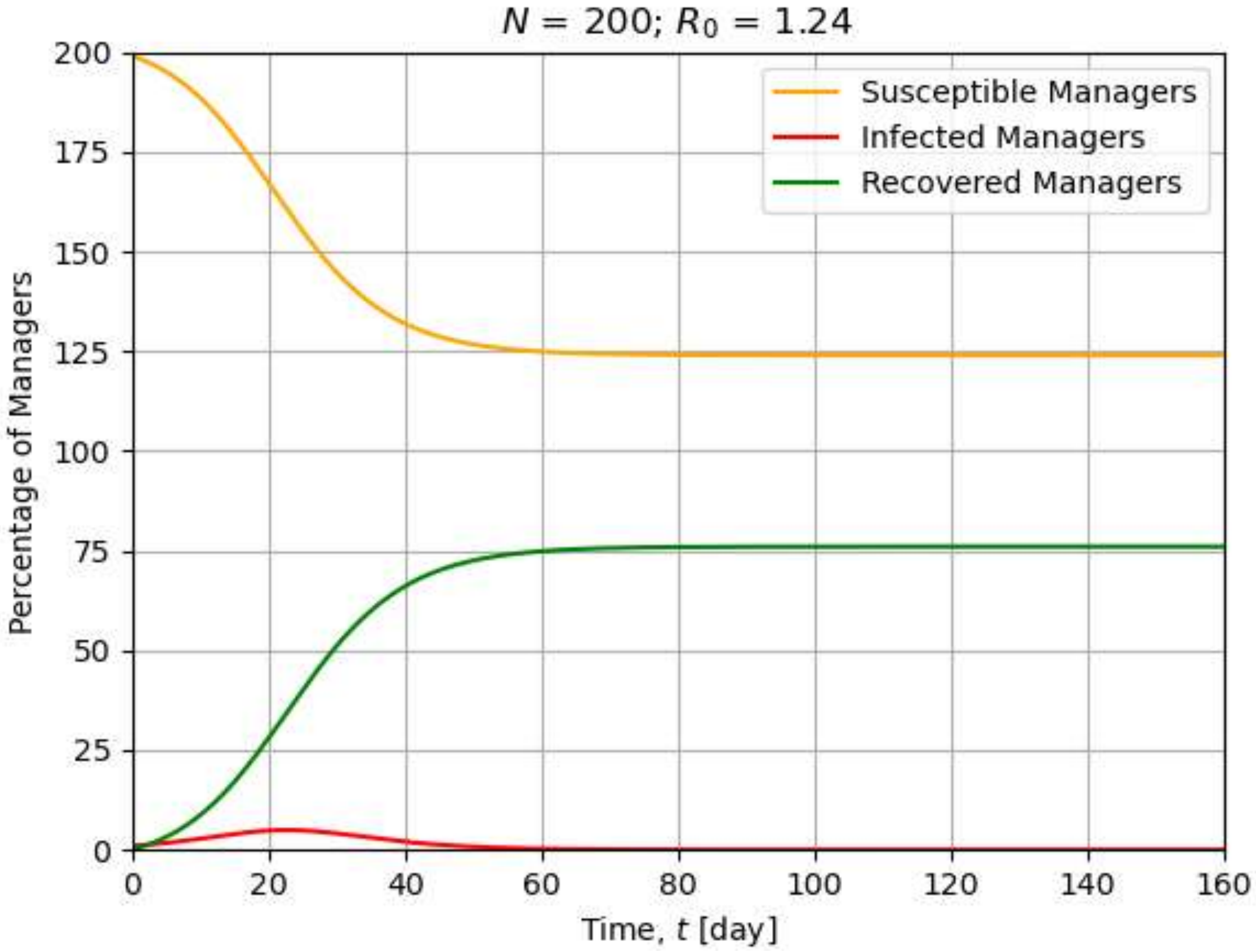

**Comments:** Ideally, information, like any disease, is spread from one participant to the other through various media and across different market sentiments. The evolution of information inception and flows (measured in terms of participants exposed) is equally crucial in financial markets. The results show that as the resistance to market flows increases, the number of susceptible and infected participants decreases significantly over time, and this is true otherwise for the recovered ones. In financial terms, traders can only get affected by market information internally if they pay attention and follow the news; otherwise, only the external (systematic) effects affect them in the long run.

## 4. Conclusions

This paper examines if there exists any interrelations among trades of any given funds in South Africa. We hypothesized that money managers and prospective investors exchange information and ideas about assets (funds) directly through word-of-mouth communication. We base our analysis on the community detection hypothesis, where we aimed to identify social networks and graphs that exist within capital markets. We analyzed the effects of information exchange on funds composition and behavior by considering 200 top mutual funds in South Africa. An Epidemiology approach has been used with some Markovian-based network models fused with statistical analysis. Our primary findings are easily summarized.

We first found that trades of any given fund type from any manager are sensitive to any information from any other manager or industry. We find that the sensitiveness is more on intra-city trades than inter-city trades. Hence, we find geographical factors as one of the critical factors in information transmission in the funds industry. Additionally, information can be transferred by word of mouth, just like contagious diseases in epidemiology. Our Epi-Finance model provided supportive results on this. We found that fund managers get manipulated by information transmitted within markets or recover from manipulation over time. If the information is negative, they can lose their investment confidence, and if they overcome, they become resistant in the long run due to confidence and market stabilization.

On the other hand, given positive information, managers can boost their portfolio sizes and returns, but this usually occurs for short periods. Based on gender and race, black managers are underrepresented in the South African funds industry but can perform equally well as whites. On the other hand, female managers can also manage successful funds. This study is part of the recent interdisciplinary studies between finance and science.